\RequirePackage{lineno}
\documentclass[%
 reprint,
twocolumn,
showpacs,preprintnumbers,showkeys,
 amsmath,amssymb,
 aps,
prc,
floatfix,
]{revtex4}

\usepackage{graphicx}
\usepackage{dcolumn}
\usepackage{bm}
\usepackage{amssymb}
\usepackage{float}
\usepackage[caption = false]{subfig}
\usepackage{hyperref}

\begin{document}    

\title{Investigating jet-medium interaction using charge dependent three-particle correlation in high multiplicity p+Pb collisions at $\sqrt{s_{NN} }$ = 5.02 TeV}
\author{Debojit Sarkar}
\email{debojit03564@gmail.com}

\affiliation{Bose Institute, CAPSS, Block EN, Sector V, Kolkata-700091, India}
\affiliation{Laboratori Nazionali di Frascati, INFN, Frascati, Italy}




\begin{abstract}

Although many observations in high multiplicity p+Pb collisions at the LHC indicate striking similarity to heavy ion collisions, no evidence of jet-medium interactions has been observed till date. We study the effect of  jet-medium interaction on the charge dependent three-particle correlation in high multiplicity p+Pb collisions  at $\sqrt{s_{NN} }$ = 5.02 TeV using EPOS 3 event generator. The short range component of the three-particle correlation is dominated by the jet fragmentation and has a unique charge dependence as observed in the minimum bias p+Au collisions at the RHIC energy. In EPOS 3,  similar pattern has  been observed in the lower multiplicity classes of p+Pb collisions where jet fragmentation plays the dominant role. Interestingly, the charge dependence of three-particle correlation gets diminished in the higher multiplicity classes of p+Pb collisions where the jet-medium interactions as implemented in EPOS 3 plays an important role. The current study can, therefore, provide possible ways to investigate the jet-medium interaction in high multiplicity classes of small collision systems at the LHC energies. 

\end{abstract}




\maketitle

\section{Introduction}
The observation of ridge \cite{CMS_pp_Ridge,alice_pPb_double_ridge} and other collective effects in high multiplicity p+p and p+Pb collisions at the LHC energies \cite{CMS_pp_v2,pPb_mass_ordering,p_pi_enhancement_pPb,CMSv2paper} are often attributed to the formation of a deconfined state of quarks and gluons similar to what is expected in the heavy ion collisions. Several experimental observations, such as,
 mass-ordering of the elliptic flow coefficient ($v_{2}$) of the identified particles  \cite {pPb_mass_ordering}, baryon-to-meson enhancement at intermediate $p_{T}$, \cite {p_pi_enhancement_pPb} and constituent quark-scaling of $v_{2}$ \cite {CMSv2paper} also suggest
similarity between the systems formed in small and heavy-ion collisions. Hydrodynamic models\cite {Shuryak_radialflow,flow_pp_victor} can explain several experimental observations in high multiplicity p+p and p+Pb collisions at LHC energies \cite {bozek_pPb,epos_ridgein_pp, epos_massordering_flow_pPb,epos_radialflow_spectra_pPb}. For example, EPOS 3 model which includes event by event 3+1 D hydrodynamic evolution \cite{epos_radialflow_spectra_pPb,epos_model_descrip} can reasonably explain the systematics of ridge \cite {epos_ridgein_pp},  mass ordering of the elliptic flow co-efficients ($v_{2}$) of identified particles \cite {epos_massordering_flow_pPb}, mass dependence of the hardening of spectra with multiplicity etc in p+p and p+Pb collisions at the LHC energies \cite {epos_radialflow_spectra_pPb,alice_pPb_radialflow,p_pi_enhancement_pPb,alice_protontopion_SQM2016,
pp_pPb_meanpt_ALICE}. These observations points to the possibility of formation of a medium describable by fluid dynamics. However, unlike heavy ion collisions, another compelling evidence of deconfined medium formation namely jet-quenching or jet-medium interaction has not been observed in the high multiplicity events of small collision systems.

Jets produced in ultra-relativistic heavy ion collisions interact with the hot and dense medium created in such collisions and  lose energy during the evolution of the system \cite{Bjorken}. This effect is referred to as jet-quenching, jet-energy loss or jet-medium interaction. This phenomena is not unique to the high energy jets as the low energy jets or mini-jets also interact with the medium and loose energy while traversing through the medium. An immediate consequence is the suppression of the invariant yields of particles when compared to the same for scaled p-p collisions at the same collision energy, referred to as the nuclear modification factor ($R_{AA}$). 
Measurements at RHIC and LHC  energies confirm medium induced modifications of particle spectra ($R_{AA} < 1$) in heavy ion collisions providing a clean evidence of jet-medium interactions \cite{Aamodt:2010jd,str}. In addition, experimental observations such as an increase in the fraction of jet pairs with largely unbalanced transverse momenta \cite{dijet_asym_1,dijet_asym_2,Krofcheck:2013dua}  and the suppression of back to back di-hadron correlation yields in the most central A+A collisions compared to the minimum bias p+p collisions \cite{dihadron_mod_1,dihadron_mod_2}  are also consistent with the notion of in-medium jet energy loss. 

However, in p+Pb collisions at the LHC energy where many features of the data indicate collectivity similar to heavy ion collisions, no evidence of suppression of hadron-yields has been observed for particles having $p_{T} >$ 10 GeV/c \cite{RpPb_1,RpPb_2}. The system size estimated from the interferometric two-particle correlations show that in the highest multiplicity p+Pb collisions, the active medium size  and the charged particle multiplicity are much less than the same in central Pb+Pb collisions \cite{systemsize_femto}. As a consequence, the possibility of interactions between the high energy jets and the medium is negligible \cite{konard}. However, as discussed in \cite{konard}, the lower energy jets (mini-jets) have a larger probability of interactions with the smaller size medium possibly formed in high multiplicity p+Pb collisions. The hadrons produced from these mini-jets populate the low $p_{T}$ part of the particle spectra and carry the signature of the jet-medium interactions. By far no convincing observable has been proposed to investigate the jet-medium interaction in high multiplicity p+Pb collisions using the low and intermediate $p_{T}$ ($p_{T} < 10$ GeV/c) particles. Experimental measurements such as medium induced modification of spectra and two particle correlations do not show any sign of dijet quenching in the small systems at the LHC energies. It is expected that any possible jet-medium interaction  in the low $p_{T}$ regime of the high multiplicity p+Pb collisions is less significant compared to the same in central heavy ion collisions and the conventional observables (spectra, di-hadron correlations etc) may not be sensitive enough to disentangle the effect of jet-medium interaction from the other low $p_{T}$ effects.

The goal of this work is to find a observable sensitive to such effects. For this, we qualitatively investigate the effect of the jet-medium interaction on the charge dependent three-particle correlation in the low $p_{T}$ ($p_{T} <$ 10 GeV/c) region of high multiplicity p+Pb collisions. The three-particle correlation has been extensively measured in small collision systems as well as in heavy ion collisions at RHIC \cite{CME_STAR,CME_STAR_3,CME_STAR_4,CME_STAR_5} and LHC \cite{CME_CMS_1,CME_CMS_2,CME_ALICE_1,CME_ALICE_rihan} energies.  The similarity in the results between p+Pb and Pb+Pb data constraints the interpretation of the observed charge-dependent correlations in heavy ion collisions as a consequence of the Chiral Magnetic effect (CME) \cite{voloshin_CME,CME_CMS_1,CME_CMS_2}. In this paper we demonstrate that the sensitivity of the charge dependent three-particle correlation towards jet-fluid interaction can be used as an useful tool to investigate the possible jet-medium interaction in high multiplicity p+Pb collisions at the LHC energies. We study the quantity $\gamma^{a,b}=C_{112}=\langle \cos(\phi_1^{a}+\phi_2^{b}-2\phi_3)$ with different charge combinations $a,b=+-,++,--$  for 0-10\% and 60-100\% event classes of p+Pb collisions at 5.02 TeV using EPOS 3 event generator with Jet-medium interaction ON (default) and OFF. The $\phi_1$, $\phi_2$ and $\phi_3$ are the azimuthal angles of the particles having 0.5  $<p_{T}< $10.0 GeV/c within $|\eta| < 0.8$. We demonstrate that the short range components of the three-particle correlation is dominated by the charge dependent two-particle correlations (jet fragmentation, resonance decays etc) and it's evolution with multiplicity can shed light on the 
possible jet-medium interaction in the higher multiplicity classes of p-Pb collisions at the LHC energies. In addition, this study will also provide insights on the charge dependent background correlations relevant for CME search in relativistic heavy ion collisions \cite{voloshin_CME}.


\section{The EPOS 3 event generator}

The EPOS 3 model includes a 3+1 D hydrodynamic evolution of bulk matter, jet production and the jet-fluid interaction \cite {epos_ridgein_pp,epos_massordering_flow_pPb,epos_radialflow_spectra_pPb,epos_model_descrip}  \cite {epos_core_corona_sep,trigdilution_EPOS_pPb,epos_PbPb_lambdakshort_enhance,ridge_EPOS_pPb_jetmedium_int}.  After the initial partonic scatterings, the final state partonic system consists of mainly longitudinal strings carrying transverse momentum of the hard scattered partons in the transverse direction. The energy loss scheme of the string segments and their positions after initial multiple scatterings (i.e inside the dense matter or at the surface) decide whether a particular string segment becomes a part of the ``core" or escapes the system to constitute the ``corona". In general, the low momentum strings in the high density core loose their individual identity  and undergo hydrodynamical evolution to form the bulk part of the system. Whereas, the highly energetic strings in the low density corona region eventually expand and finally break via the production of quark-antiquark or diquark-antidiquark pairs following Schwinger mechanism to produce jets.  However, the intermediate $p_{T}$ string segments have a formation time such that they break down inside the core and  the string breaking (i.e fragmentation) is influenced by the flowing bulk matter \cite{EPOS_main}. These segments may pick up quarks or antiquarks needed for the string breaking from the fluid (bulk)  \cite {epos_PbPb_lambdakshort_enhance} rather than generating through the Schwinger mechanism in vacuum. The produced jet-hadrons are composed of a string segment originating from the initial hard process and di(quarks) from the fluid, and therefore, carry fluid properties. This effect is more pronounced in the higher multiplicity classes where the transverse size of the system is higher, increasing the probability of fluid-jet interaction  \cite {epos_PbPb_lambdakshort_enhance}. There is also an option to switch off the default interaction between jet (corona) and medium (core) in EPOS 3. In that case core and corona hadronizes separately and the string fragmentation is not influenced by the flowing bulk matter. However, the interaction between the jets and flowing bulk matter has been found to be essential to describe the inclusive baryon-to-meson enhancement at intermediate $p_{T}$  in Pb-Pb collisions at 2.76 TeV. In addition, this jet-fluid interaction in EPOS has also been found to contribute to the ridge  in the higher multiplicity classes of p+Pb collisions  \cite{ridge_EPOS_pPb_jetmedium_int}. But, experimentally it is very challenging to disentangle the different sources (e.g hydro, jet-fluid interaction) contributing to the ridge in high multiplicity p+Pb collisions, and therefore, studying jet-medium interaction using di-hadron correlation becomes difficult.\\ 

We measure the charge dependent three-particle correlation in EPOS 3 with the jet-medium interaction ON(default) and OFF in 0-10\% and 60-100\% event classes. We demonstrate that the modification of the charge dependence  of 3 particle correlations at the higher multiplicity classes compared to the lower ones may provide important insight on the possible jet-fluid interaction in small collision systems.

\begin{figure}[htb!]

\includegraphics[scale=0.40]{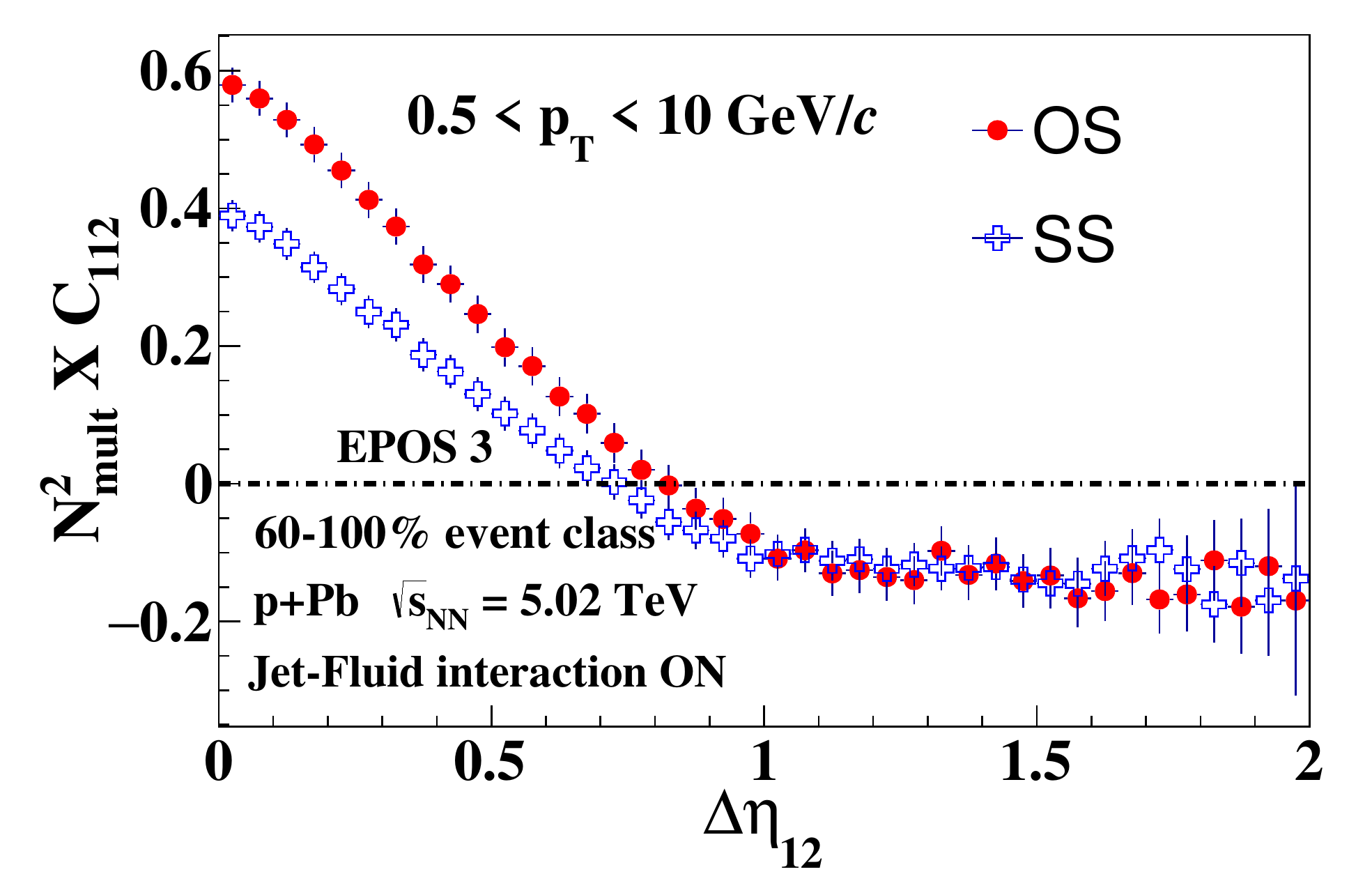}

\caption{[Color online] The $\Delta\eta_{12}$ ($\!\eta_1\!-\!\eta_2$ )  dependence of $C_{112}$ scaled by $N_{\rm mult}^{2}$ for the same and opposite sign pairs between 1st and 2nd particle in the 60-100\% event class of p+Pb collisions in EPOS 3 with the default jet-fluid interaction ON. }
\label{partondist_XY}
\end{figure}

\section{Results and Discussion}


We measure the relative rapidity $\Delta\eta_{12}=\!\eta_1\!-\!\eta_2$ dependence of the charge dependent (same and opposite sign pairs between 1st and 2nd particle) three-particle correlator $C_{112} (\eta_1-\eta_2)=\langle \cos(\phi_1 (\eta_1)+\phi_2 (\eta_2)-2\phi_3)\rangle$ in the different multiplicity classes of p+Pb collisions. In Fig 1 we show $C_{112} (\eta_1-\eta_2)$ for the 60-100\% event class of p+Pb collisions at 5.02 TeV using EPOS 3 event generator with the default jet-fluid interaction ON. To take into account the natural dilution of correlation with multiplicity \cite{3part_corr_STAR}, $C_{112}$ is scaled by $N_{\rm mult}^{2}$ where $N_{\rm mult}$ is the number of charged hadrons with 0.5  $<p_{T}< $10.0 GeV/c within $|\eta| < 0.8$. 
The multiplicity classes have been estimated based on the total amount of charged particles produced (with $p_{T}$ $>$0.05 GeV/c) within 2.8  $<\eta<$5.1. This corresponds to the acceptance range of ALICE VZERO-A detector in the Pb going direction in case of p+Pb collisions and used for multiplicity class determination by the ALICE collaboration \cite {alice_pPb_double_ridge},\cite {pPb_mass_ordering}.\\

In Fig 1 we see that $C_{112}$ remains positive at small $\Delta\eta$ and changes sign at large $\Delta\eta$. Different underlying phenomena contribute towards the $\Delta\eta$-dependence of $C_{112}(\Delta\eta)$~\cite{Adamczyk:2017hdl}. The charge dependence of the three particle correlator with the ordering $C_{112}^{{opposite-sign}} > C_{112}^{same-sign} > 0$ at smaller $\Delta\eta$ is a characteristic feature of the string fragmentation and has already been observed in the minimum bias p+Au collisions at 200 GeV \cite{CME_STAR}. The  effects of initial state geometry, hydrodynamic response and momentum conservation lead to negative values of $C_{112}$ as shown in previous studies \cite{Bzdak:2010fd,Adamczyk:2017hdl,Teaney:2010vd}. Such effects are responsible for the observation of $C_{112}\!<\!0$ at larger $\Delta\eta$ as shown in Fig 1. It is important to note that at large $\Delta\eta$ ($|\Delta\eta| > $ 1) where the effect of initial geometry and hydrodynamic response play an important role, the OS and SS correlation converge to similar value. We investigate whether the charge dependence of $C_{112}$ at smaller $\Delta\eta$ gets modified with multiplicity due to possible jet-medium interaction at higher multiplicity classes of p+Pb collisions.


In this analysis, the three particle correlator is constructed using the low $p_{T}$ particles (0.5  $<p{_T}<$ 10.0 GeV/$\it{c}$) to take into account the particles originating from the fragmentation of lower energy jets which have larger probability of fragmentation within the freeze-out hypersurface in EPOS 3 \cite{ridge_EPOS_pPb_jetmedium_int,epos_PbPb_lambdakshort_enhance}. In Fig 2(a)  the relative rapidity $\Delta\eta_{12}=\!\eta_1\!-\!\eta_2$ dependence of the charge dependent three-particle correlator $C_{112}$ for the 0-10\% event class of p+Pb collisons at 5.02 TeV is shown with the jet-fluid interaction ON (default). Interestingly, the charge dependence of $C_{112}$ at smaller $\Delta\eta$ gets diminished in the higher multiplicity classes of p+Pb collisions. The reason for diminishing of the charge dependence of $C_{112}$ with increase in multiplicity can be understood as follows. 

\begin{figure}
a)

\includegraphics[scale=0.40]{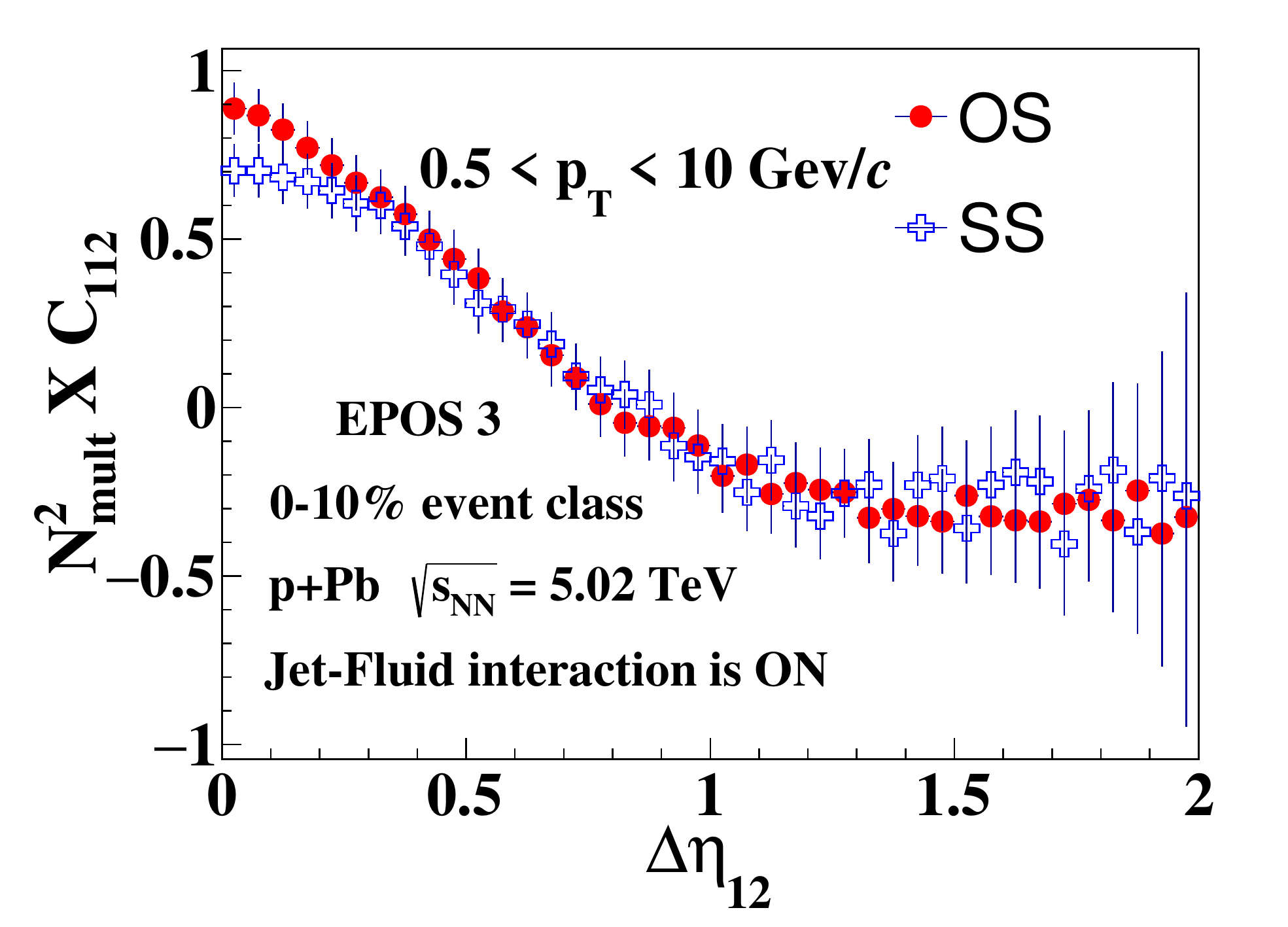}

b)

\includegraphics[scale=0.43]{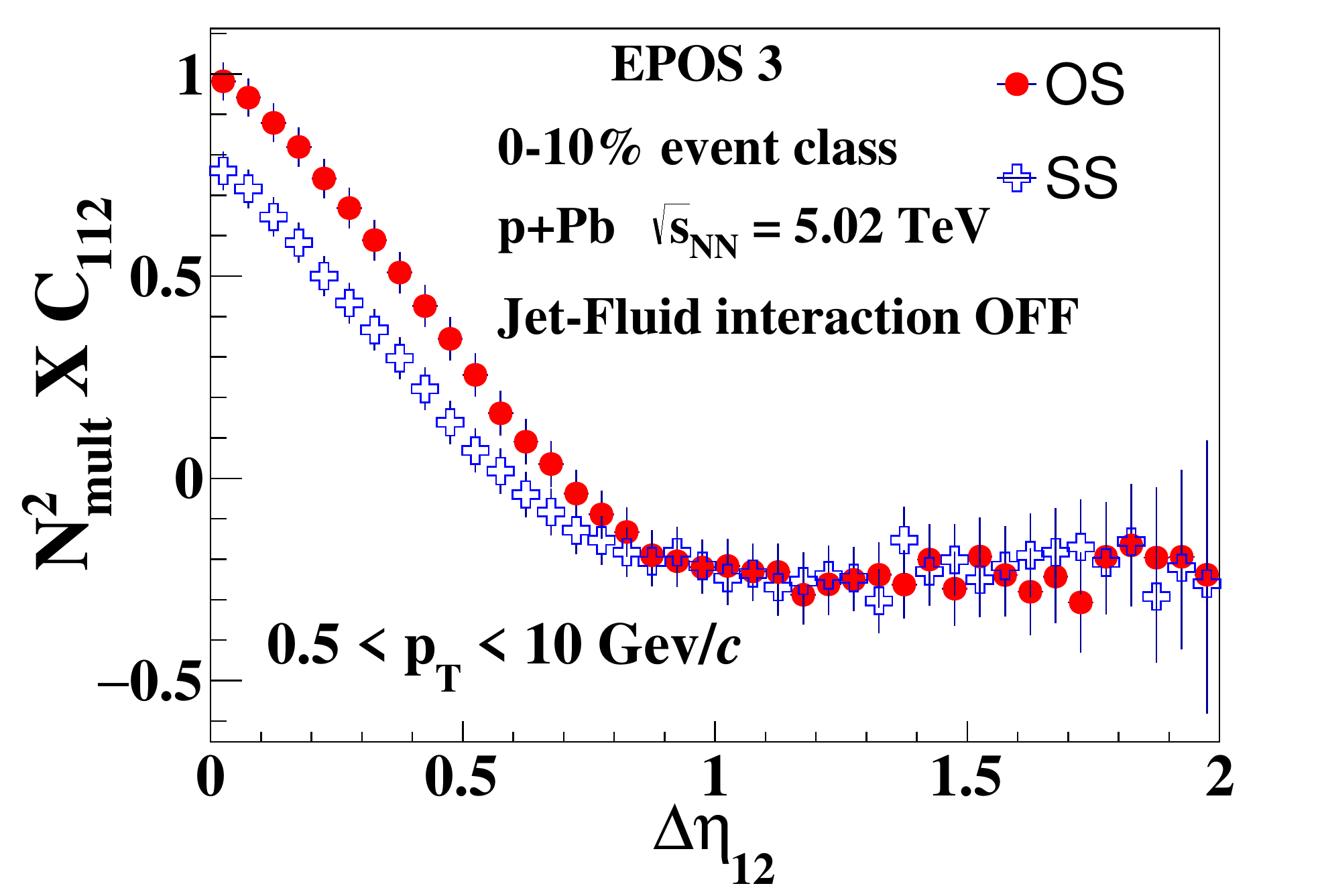}

\caption{[Color online] a) The $\Delta\eta_{12}$ ($\!\eta_1\!-\!\eta_2$ )  dependence of $C_{112}$ scaled by $N_{\rm mult}^{2}$ for the same and opposite sign pairs between 1st and 2nd particle in the 0-10\% event class of p+Pb collisions in EPOS 3 with the default jet-fluid interaction ON.  b) Same as Fig (a) but with the  jet-fluid interaction OFF. }
\label{partondist_XY}
\end{figure}

Aforementioned, with increase in multiplicity, the size of the core and the probability of fluid-jet interaction increases. The jet hadrons (corona) produced via string breaking inside or at the surface of the bulk  use partons from the bulk (core) and therefore carry fluid properties \cite {ridge_EPOS_pPb_jetmedium_int,epos_PbPb_lambdakshort_enhance}. As the number of jet hadrons produced from jet-fluid interaction and therefore carrying the fluid properties increases with multiplicity/system size \cite {epos_PbPb_lambdakshort_enhance}, this will affect the charge dependence of the $C_{112}$ at smaller $\Delta\eta$. The jet hadrons carrying fluid partons are expected to show similar pattern as observed in the long range of $C_{112}$ where the hydrodynamic response plays a significant role. 
For a better understanding of the effect of jet-fluid interaction on the short range of $C_{112}$, we repeat the analysis with jet-fluid interaction turned OFF for the 0-10\% event class of p+Pb collisions as shown in Fig 2(b). A charge dependence of $C_{112}$ ($C_{112}^{{opposite-sign}} > C_{112}^{same-sign}$) at smaller $\Delta\eta$ is observed in Fig 2(b) which is qualitatively similar to the one observed in 60-100\% event class of p+Pb collisions where jet fragmentation is the dominant mechanism of particle production. This indicates that the diminishing of charge dependence of $C_{112}$ at small $\Delta\eta$  in the high multiplicity p+Pb collisions has contribution from the jet-fluid interaction as implemented in EPOS 3.

Let us discuss if this observable is a robust one and if there is any other phenomenon that can mimic the effect of jet-medium interaction as shown in Fig.1. and Fig 2. The charge dependence of the three-particle correlation can arise from different origins such as CME \cite{voloshin_CME}, local charge conservation \cite{CME_bkg_2}, flowing   resonances \cite{voloshin_CME,CME_bkg_4}, di-jet fragmentation \cite{CME_STAR} and initial state correlations\cite{CME_initial_state}. As CME is driven by the magnetic filed, this is not a possible explanation of the charge dependence of  three-particle correlation in p+Pb collisions. Local charge conservation coupled with anisotropic flow, driven by initial state geometry, can be a possible explanation of such effect but such presumption is still under debate as it can't explain all aspects of the experimentally observed three-particle correlations at RHIC and LHC energies \cite{CME_bkg_1,CME_bkg_2,CME_bkg_3,CME_bkg_4}. The charge independent effects (e.g momentum conservation) are cancelled out during ($\Delta C_{112} = C_{112}^{OS}$ - $C_{112}^{SS}$) and the contribution from local charge conservation and resonance decay are suppressed by the number of particles resulting in a dependence of $\Delta C_{1,1,2} \approx v_2^2/N_{\rm mult}$ \cite{voloshin_CME} where $v_2^2 = <cos(2 \phi_1 -2 \phi_2)>$ is the measured elliptic anisotropy of the system. Therefore, one expects ($\Delta C_{112}*N_{\rm mult}) / v_2^2$ to be the independent of $N_{\rm mult}$. In case of p+Pb collisions, $v_{2}$ approximately varies as 1/ $\sqrt{N_{\rm mult}}$ as the initial spatial anisotropy has a dominant contribution from random fluctuations in the positions of the participating nucleons, non-flow from di-jets and other initial state momentum space correlations\cite{v2_vs_N_1,v2_vs_N_2,3part_corr_STAR}. As a result, one would naively expect that $\Delta C_{112} * N_{\rm mult}^{2}$ would be independent of $N_{\rm mult}$. However, we observe that the $\Delta C_{112} *N_{\rm mult}^{2}$ decreases with increase in multiplicity ($N_{\rm mult}$) and the jet-medium interaction has an effect on the multiplicity evolution of $\Delta C_{112} *N_{\rm mult}^{2}$. Most importantly, $\Delta C_{112} *N_{\rm mult}^{2}$ shows similar pattern in the 0-10\% and 60-100\% event classes of p+Pb collisions when the jet-medium interaction is switched off. Therefore, we argue that this observable can be used as an useful tool to investigate the possible jet-medium interactions in high multiplicity classes of small collision systems.

It should be noted that, till date, most of the experimental measurements at the RHIC and the LHC energies concentrate on the multiplicity evolution of $\gamma_{112} = C_{112}/v_{2}$ to search for CME in relativistic heavy ion collisions. Therefore, a direct data-model comparisons for $\Delta C_{112} *N_{\rm mult}^{2}$ is not possible at this moment. But, from previous measurements it is clear that the existing event generators cannot explain the experimentally observed three-particle correlation results in heavy ion collisions as well as in small collision systems \cite{CME_bkg_1}. In case of heavy ion collisions, CME is often considered as a viable explanation for such discrepancies. However, as the effect of CME is expected to be negligible in the p+Pb collisions, the inability  of the existing event generators to properly describe the three-particle correlation results suggests that some important physics could be missing in such models \cite{CME_bkg_1}. Efforts are ongoing to better understand the background sources of correlations resembling CME like effect in the experimental measurements \cite{CME_bkg_1,CME_bkg_2}. Our study shows that the multiplicity evolution of charge dependence of $C_{112}$ can also be used as a probe to study the potential interaction between the low energy jets and the medium in high multiplicity classes of small collision systems. Further studies on the charge dependent three particle correlations with identified particles will be helpful in constraining different models aiming to explain the collective behaviors in high multiplicity small collision systems as well as shed light on the different sources of charge dependent correlations contribute as a background in the CME search activities.

\section*{Acknowledgements} 
I acknowledge fruitful discussions and suggestions from Prithwish Tribedy. I thank Klaus Werner for allowing me to use EPOS 3 for this study. I would like to acknowledge the financial support from the CBM-MUCH project grant of BI-IFCC/2016/1082(A) and thankful to Sanjay Kumar Ghosh, Subhasis Chattopadhyay and Supriya Das for their help and support. I also thank Federico Ronchetti, Alessandra Fantoni and Valeria Muccifora from INFN Frascati for their help and support. Thanks to the VECC grid computing team for their constant effort to keep the facility running and helping in EPOS data generation and data analysis.


\end{document}